# Circular Polarization and Coherent Backscattering

Adrian J. Brown,[1*]

[1]Plancius Research, Severna Park, MD, USA 21146

*Corresponding author: adrian.j.brown@nasa.gov

## Abstract

We extend the work of Mishchenko et al. (2000) regarding the exact results of the polarization effect, using the theory of Amic et al. (1997) to derive a model prediction for the polarization coherent opposition effect at small angles. Our extension is to assess the effect of circular polarized light, thus completing exact derivation of the full Müller matrix for the semi infinite slab of Rayleigh sized particles. We find the circular polarization peak is narrower than the coherent backscattering intensity peak, and weaker in intensity.

## 1. Introduction

Amic et al. [1] developed a theoretical model based on an assessment of the full Müller matrix scattering behavior of rayleight scatterers. Mishchenko et al. [2] utilized this model to assess the coherent backscatter peak and linear polarization ('negative polarization curve' [3]). Here we utilize the Amic et al. model to make an assessment of the circular polarization coherent backscatter signature.

Furthermore, we extend the results to a full Müller matrix hemispherical map of the backscattered polarized light for a Rayleigh scattering target.

We show that the circular polarization backscatter peak is we find the circular polarization peak is narrower than the intensity peak, and also a little weaker. Both of these features are stronger in in intensity than the negative linear polarization peak, however the negative linear polarization peak is wider than either the intensity or circular polarization. By presenting the relative intensities and widths of the full backscattering Müller matrix for a slab of Rayleigh particles in an exact calculation, we provide testable hypotheses for comparison with future experimental investigations of the coherent backscattering peak.

### 1.1 Definitions and Parameter Space

Amic et al. [1] developed a Milne equation based model of the coherent backscattering effect, based on the work of Tsang and Ishimaru [4] van der Mark [5] van Tiggelen [6] and Ozrin [7]. Here we give an outline of that derivation, the interested reader is referred to the original source for more details.

Using the following form of the Stokes parameters,

$$I = (I_1, I_2, I_3, I_4)^T \text{ where}$$
$$I_1 = \frac{1}{2}(I+Q) = \langle E_\parallel E_\parallel^* \rangle = |E_\theta|^2$$
$$I_2 = \frac{1}{2}(I-Q) = \langle E_\parallel E_\parallel^* \rangle = |E_\phi|^2 \quad (1)$$
$$I_3 = U = \langle E_\parallel E_\perp^* \rangle + \langle E_\perp E_\parallel^* \rangle = 2\operatorname{Re}\langle E_\theta E_\phi^* \rangle$$
$$I_4 = V = i(\langle E_\parallel E_\parallel^* \rangle + \langle E_\perp E_\perp^* \rangle) = 2\operatorname{Im}\langle E_\theta E_\phi^* \rangle$$

The mean diffuse radiation propagating in the direction $(\theta, \varphi)$ in a scattering medium at depth $\tau$ is described by the Stokes vector $\mathbf{I}(\tau, \mu, \varphi)$ with:

$$\mu = \cos\theta, \nu = \sin\theta = \sqrt{1-\mu^2} \quad (2)$$

The radiative transfer equation we shall deal with here is:

$$\mu \frac{\partial}{\partial t} \mathbf{I}(\tau, \mu, \phi) = \mathbf{\Gamma}(\tau, \mu, \phi) - \mathbf{I}(\tau, \mu, \phi) \quad (3)$$

and the source function $\mathbf{\Gamma}$ for the Rayleigh scattering phase function $\mathbf{P}(\mu, \varphi, \mu', \varphi')$ is:

$$\mathbf{\Gamma}(\tau, \mu, \phi) = \int_{-1}^{1} \frac{d\mu'}{2\mu'} \int_{0}^{2\pi} \frac{d\phi'}{2\pi'} \mathbf{P}(\mu, \phi, \mu', \phi') \mathbf{I}(\tau, \mu', \phi') \quad (4)$$

where $\mu, \varphi$ are scattering angles for incoming light, and $\mu', \varphi'$ are outgoing scattering angles. $\tau$ is the depth parameter for penetration into the scattering medium.

## 2. Formulation of Schwarzchild-Milne Equation

We simplify the full multiple scattering transfer equation for $\Gamma$ when we express it as the Schwarzchild-Milne [8], [9] equation:

$$\Gamma(\tau,\mu,\phi) = P(\mu,\phi,\mu_a,\phi_a) I_a e^{-\tau/\mu_a}$$
$$+ \int_0^\tau d\tau' \int_0^1 \frac{d\mu'}{2\mu'} \int_0^{2\pi} \frac{d\phi'}{2\pi} e^{-(\tau-\tau')/\mu'} P(\mu,\phi,\mu',\phi') \Gamma(\tau',\mu',\phi')$$
$$+ \int_\tau^\infty d\tau' \int_0^1 \frac{d\mu'}{2\mu'} \int_0^{2\pi} \frac{d\phi'}{2\pi} e^{-(\tau'-\tau)/\mu'} P(\mu,\phi,-\mu',\phi') \Gamma(\tau',-\mu',\phi') \quad (5)$$
$$+ \int_0^\infty d\tau' \int_0^1 \frac{d\mu'}{2\mu'} \int_0^{2\pi} \frac{d\phi'}{2\pi} e^{-(\tau+\tau')/\mu'} R(\mu') P(\mu,\phi,\mu',\phi') \Gamma(\tau',\mu',\phi')$$

There are four terms here - the first is due to incident light, the second and third are due to scattering from the bulk of the material (away from the boundaries) from above and below and the final term is 'skin layer' term which accounts for light scattered at depth and then reflected once at the boundary [10].

For this choice of Stokes parameters, the reflection matrix is defined as:

$$R(\mu) = \begin{bmatrix} |r_\parallel|^2 & 0 & 0 & 0 \\ 0 & |r_\perp|^2 & 0 & 0 \\ 0 & 0 & \mathrm{Re}(r_\parallel r_\perp^*) & -\mathrm{Im}(r_\parallel r_\perp^*) \\ 0 & 0 & \mathrm{Im}(r_\parallel r_\perp^*) & \mathrm{Re}(r_\parallel r_\perp^*) \end{bmatrix} \quad (6)$$

where the Fresnel reflection coefficients are

$$r_\parallel = r_\parallel(\mu) = \frac{\mu - m\sqrt{1-m^2 v^2}}{\mu + m\sqrt{1-m^2 v^2}} \text{ and } r_\perp = r_\perp(\mu) = \frac{\sqrt{1-m^2 v^2} - m\mu}{\sqrt{1-m^2 v^2} + m\mu}$$

For the purposes of this study, we neglect the 'skin layer' term and set R = 0. We are thus ignoring the possibility of reflections from the boundary back into the layer. This situation mimics the case where optical properties of the slab and the exterior environment are closely matched [10]. Further terms of multiple scattering, for example for rescattering of the light from the boundaries more than once, are also neglected in this approximation.

We can then expand the φ dependence of the intensity and phase matrix in complex trigonometric polynomials $e^{ik\phi}$.

$$I(\mu,\phi) = \sum_{k=-2}^{2} I^{(k)}(\mu) e^{ik\phi}$$
$$\Gamma(\mu,\phi) = \sum_{k=-2}^{2} \Gamma^{(k)}(\mu) e^{ik\phi} \quad (7)$$
$$P(\mu,\phi) = \sum_{k=-2}^{2} P^{(k)}(\mu) e^{ik\phi}$$

The Rayleigh phase matrices of Chandrasehkar [11] (his p. 42 eqn. 221-224) are then modified with this expansion to read:

$$P^{(0)}(\mu,\mu') = \frac{3}{4} \begin{bmatrix} 2(1-\mu^2)(1-\mu'^2)+\mu^2\mu'^2 & \mu^2 & 0 & 0 \\ \mu'^2 & 1 & 0 & 0 \\ 0 & 0 & 0 & 0 \\ 0 & 0 & 0 & 2\mu\mu' \end{bmatrix}$$

$$P^{(1)}(\mu,\mu') = P^{(-1)*}(\mu,\mu') = \frac{3}{4} vv' \begin{bmatrix} \mu\mu' & 0 & i\mu & 0 \\ 0 & 0 & 0 & 0 \\ -2i\mu' & 0 & 1 & 0 \\ 0 & 0 & 0 & 1 \end{bmatrix} \quad (8)$$

$$P^{(2)}(\mu,\mu') = P^{(-2)*}(\mu,\mu') = \frac{3}{8} \begin{bmatrix} \mu^2\mu'^2 & -\mu^2 & -i\mu^2\mu' & 0 \\ -\mu'^2 & 1 & -i\mu' & 0 \\ -2i\mu\mu'^2 & 2i\mu & 2\mu\mu' & 0 \\ 0 & 0 & 0 & 0 \end{bmatrix}$$

and the Schwarzchild-Milne equation is split into five decoupled integral equations:

$$\Gamma^{(k)}(\tau,\mu) = P^{(k)}(\mu,\mu_a) I_a e^{-ik\phi_a - \tau/\mu_a}$$
$$+ \int_0^\tau d\tau' \int_0^1 \frac{d\mu'}{2\mu'} e^{-(\tau-\tau')/\mu'} P^{(k)}(\mu,\mu') \Gamma^{(k)}(\tau',\mu')$$
$$+ \int_\tau^\infty d\tau' \int_0^1 \frac{d\mu'}{2\mu'} e^{-(\tau'-\tau)/\mu'} P^{(k)}(\mu,-\mu') \Gamma^{(k)}(\tau',-\mu') \quad (9)$$
$$+ \int_0^\infty d\tau' \int_0^1 \frac{d\mu'}{2\mu'} e^{-(\tau+\tau')/\mu'} R(\mu') P^{(k)}(\mu,\mu') \Gamma^{(k)}(\tau',\mu')$$

This is a considerable simplification, because we have eliminated the φ' dependence from our equations and we may now attack each of the five equations separately.

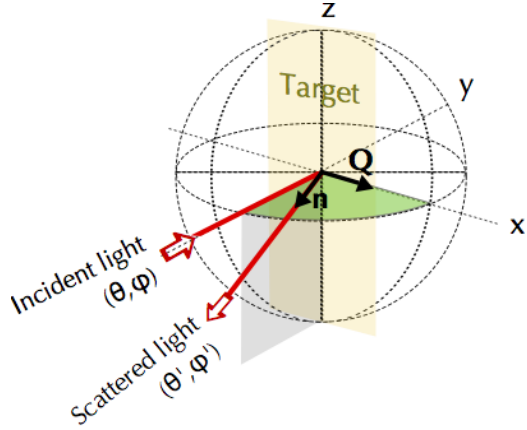

Fig. 1 – Definition of scattering planes for incident and scattered light used in this study.

### 2.1 Description of Enhanced Backscattering Cone

To make progress in our examination of the nature of the enhanced backscattering cone, it is useful to define the dimensionless transverse wavevector **Q**, which we define in terms of the mean free path $l$,

$$\mathbf{Q} = \mathbf{q} l \quad (10)$$

with a magnitude

$$Q = ql = kl\theta = k_1 l \theta_1 \quad (11)$$

and here we assume **Q** is parallel to the x-axis (Figure 1). We now insert **Q** as a small perturbation in the exponent of the leading term in equation (9) by making the canonical replacement:

$$e^{-\tau/\mu'} \to e^{-\tau/\mu' + i\mathbf{Q}\cdot\mathbf{n}\tau/\mu'} = e^{-(1-i\mathbf{Q}\cdot\mathbf{n})\tau/\mu'} = e^{-(1-iQv'\cos\phi')\tau/\mu'} \quad (12)$$

The last equality comes because **n** is the unit vector in the direction $(\theta',\phi')$. When this replacement is made in equation (9) and using the following properties of Bessel functions:

$$i^n J_n(z) = \int_0^{2\pi} \frac{\partial \phi}{2\pi} e^{iz\cos\phi - in\phi} \text{ and}$$
$$J_{-n}(z) = J_n(-z) = (-1)^n J_n(z) \quad (13)$$

we obtain, upon expanding in complex trigonometric polynomials $e^{ik\phi}$:

$$\Gamma^{(k)}(\tau,\mu) = P^{(k)}(\mu,\mu_a) I_a e^{-ik\phi_a - \tau/\mu_a}$$
$$+ \int_0^\tau d\tau' \int_0^1 \frac{d\mu'}{2\mu'} e^{-(\tau-\tau')/\mu'} P^{(k)}(\mu,\mu') \sum_{j=-2}^{2} i^{k-j} J_{k-j}(Q(\tau-\tau')v'/\mu') \Gamma^{(k)}(\tau',\mu')$$
$$+ \int_\tau^\infty d\tau' \int_0^1 \frac{d\mu'}{2\mu'} e^{-(\tau'-\tau)/\mu'} P^{(k)}(\mu,-\mu') \sum_{j=-2}^{2} i^{k-j} J_{k-j}(Q(\tau'-\tau)v'/\mu') \Gamma^{(k)}(\tau',-\mu') \quad (14)$$
$$+ \int_0^\infty d\tau' \int_0^1 \frac{d\mu'}{2\mu'} e^{-(\tau+\tau')/\mu'} R(\mu') P^{(k)}(\mu,\mu') \sum_{j=-2}^{2} i^{k-j} J_{k-j}(Q(\tau+\tau')v'/\mu') \Gamma^{(k)}(\tau',\mu')$$

### 2.2 Reflected Intensity and the Bistatic matrix equation

We now look closely at the total reflected intensity $A(Q)$ near the backscattering direction, which is defined as [4]:

$$A(Q) = A^L + A^C(Q) - A^{SS} = \frac{4}{\pi*(m+1)^4} \langle I_b | L \cdot \gamma^L + \gamma^C - \gamma^{SS} | I_a \rangle \quad (15)$$

Where $I_a$ is the incoming radiation and $I_b$ is the outgoing and $L$ is the diagonal rotation matrix $L = \text{diag}\{1, 1, -0.5, 0.5\}$.

The reflected intensity is the sum of three parts, $\gamma^L$, $\gamma^C$ and $\gamma^{SS}$. These are the contributions from the ladder diagrams, maximally crossed diagrams and single scattering (also known as Diffuson and Cooperon contributions [12]). The Diffuson or ladder bistatic matrix equation at normal incidence is as follows:

$$\gamma^L = \gamma(\mu_a=1, \phi_a=0, \mu_b=1, \phi_b=0) = \begin{bmatrix} \gamma_{11} & \gamma_{12} & 0 & 0 \\ \gamma_{12} & \gamma_{11} & 0 & 0 \\ 0 & 0 & \gamma_{12} - \gamma_{11} & 0 \\ 0 & 0 & 0 & \gamma_{44} \end{bmatrix} \quad (16)$$

and the maximally crossed diagrams bistatic matrix equation at normal incidence is:

$$\gamma^C(Q) = \begin{bmatrix} \gamma_{11} & \tilde{\gamma}_{12} & 0 & 0 \\ \tilde{\gamma}_{12} & \gamma_{11} & 0 & 0 \\ 0 & 0 & \tilde{\gamma}_{33} & 0 \\ 0 & 0 & 0 & \tilde{\gamma}_{44} \end{bmatrix}$$

with
$$\tilde{\gamma}_{12} = \tilde{\gamma}_{12}(Q) = \frac{1}{2}(\gamma_{44} - \gamma_{33}) \quad (17)$$
$$\tilde{\gamma}_{33} = \tilde{\gamma}_{33}(Q) = \frac{1}{2}(\gamma_{44} + \gamma_{33}) - \gamma_{12}$$
$$\tilde{\gamma}_{44} = \tilde{\gamma}_{44}(Q) = \frac{1}{2}(\gamma_{44} + \gamma_{33}) + \gamma_{12}$$

the final bistatic matrix accounts for single scattering events, which are their own time-reversed counterparts so they appear in both $\gamma^L$, $\gamma^C$ and must not be double counted:

$$\gamma^{SS}(Q) = \frac{3}{4} \begin{bmatrix} 1 & 0 & 0 & 0 \\ 0 & 1 & 0 & 0 \\ 0 & 0 & -1 & 0 \\ 0 & 0 & 0 & -1 \end{bmatrix} \quad (18)$$

### 2.3 Parameterization of Schwarzchild-Milne equation

We now wish to solve for the source function $\Gamma$ in equation (14). To bring this to fruition, it is useful to introduce the following parameterization for the five expansions of $\Gamma$, which is derived for k=0 in Appendix A:

$$\boldsymbol{\Gamma}^{(0)}(\tau,\mu) = \begin{bmatrix} A(\tau) + B(\tau)(1-\mu^2) \\ A(\tau) \\ 0 \\ C(\tau)\mu \end{bmatrix}$$

$$\boldsymbol{\Gamma}^{(1)}(\tau,\mu) = -\boldsymbol{\Gamma}^{(-1)*}(\tau,\mu) = v \begin{bmatrix} (iD(\tau) - E(\tau))\mu \\ 0 \\ D(\tau) + iE(\tau) \\ iF(\tau) \end{bmatrix} \quad (19)$$

$$\boldsymbol{\Gamma}^{(2)}(\tau,\mu) = -\boldsymbol{\Gamma}^{(-2)*}(\tau,\mu) = (G(\tau) + iH(\tau)) \begin{bmatrix} \mu^2 \\ -1 \\ -2i\mu \\ 0 \end{bmatrix}$$

where capital latin letters are defined as below:

$$A = \frac{3}{4}((I_1 + I_2)e^{-\tau} + (M_{00} + M_{02})*A + (M_{02} - M_{04})*B - 2M_{13}*D + 2(M_{20} + M_{22})*G)$$
$$B = \frac{3}{4}(-(I_1 + I_2)e^{-\tau} + (M_{00} + 3M_{02})*A + (2M_{00} - 5M_{02} + 3M_{04})*B + 2(-2M_{11} + 3M_{13})*D - 2(\;$$
$$D = \frac{3}{4}(2M_{11}*A + 2(M_{11} - M_{13})*B + (M_{00} + M_{02} - 2M_{04} + M_{20} - 2M_{22})*D + 2(M_{11} + M_{13} + M_{31})\;$$
$$G = \frac{3}{8}((I_1 - I_2)e^{-\tau} + M_{20}*A - M_{22}*B - (M_{11} + M_{13} + M_{31})*D + (M_{00} + 2M_{02} + M_{04} + M_{40})*G) \quad (20)$$
$$C = \frac{3}{2}(I_4 e^{-\tau} + M_{02}*C - 2M_{11}*F)$$
$$F = \frac{3}{4}(M_{11}*C + M_{11}*C + (M_{00} - M_{02} - M_{20})*F)$$
$$E = \frac{3}{4}((M_{00} + M_{02} - 2M_{04} - M_{20} + 2M_{22})*E + 2(M_{11} + M_{13} - M_{31})*H)$$
$$H = \frac{3}{8}(I_3 e^{-\tau} + (-M_{11} - M_{13} + M_{31})*E + (M_{00} + 2M_{02} + M_{04} - M_{40})*H)$$

with the Milne kernel for these equations defined as (note this equation has a small correction to the order of the subindices of $M$ compared to equation (3.48) of Amic. et al – they are interchanged relative to Amic et al to make them correct):

$$M_{qp}(Q,\tau) = \int_0^1 (\frac{\partial \mu}{2\mu}) \mu^p v^q e^{\frac{-|\tau|}{\mu}} J_q(Qv\frac{|\tau|}{\mu}) \quad (21)$$

and the convolution operator * is defined as:

$$(M*A)(\tau) = \int_0^{+\infty} M(\tau - \tau') A(\tau') d\tau' \quad (22)$$

## 3. Solutions of Schwarzschild-Milne equation

We have thus learned that the decoupled Schwarzschild-Milne equations (9) have the form of a convolution equation, and this is suggestive of a potential solution deploying a Laplace transform and Wiener-Hopf attack in order to solve the resultant singular integrals. The Laplace transform is defined as:

$$a(s) = \int_0^{+\infty} A(\tau) e^{s\tau} d\tau \quad (23)$$

We therefore first take the Laplace transform of equation (21) and using the relation for the Laplace transform of a Bessel function of the $n$th order:

$$\int_0^\infty J_n(b\tau) e^{(-\tau a)} d\tau = \frac{1}{2\pi} \int_{-\pi}^{\pi} \frac{e^{in\theta}}{a + ib\sin\theta} d\theta \quad (24)$$

we obtain, setting $a = \tau/|\mu|$ and $b = Qv\tau/|\mu|$, for example, for the $_{pq=00}$ element:

$$m_{00}(Q,s) = \int_{-1}^{1} \frac{d\mu}{2} \int_0^{2\pi} \frac{d\phi}{2\pi} \int_{-\infty}^{+\infty} \frac{d\tau}{|\mu|} \Theta(\tau\mu) \exp\left(s\tau - \frac{\tau}{\mu} + iQv\frac{\tau}{\mu}\sin\phi\right) \quad (25)$$

Performing the $\tau$ integral, and transforming to polar coordinates such that:

$$\begin{aligned} X &= v\cos\theta = \sin\theta\cos\phi \\ Y &= v\sin\theta = \sin\theta\sin\phi \\ Z &= \mu = \cos\theta \end{aligned} \quad (26)$$

yields specifically for $m_{00}$:

$$m_{00}(Q,s) = \int \frac{d\Omega}{4\pi} \frac{1}{1 - s\cos\theta - iQ\sin\theta\cos\phi} \quad (27)$$

and we can finally apply a transform into 'sigma space' like this:

$$\sigma = \sqrt{(s^2 - Q^2)} \quad (28)$$

to obtain:

$$\begin{aligned}
m_0(\sigma) &= \frac{1}{2\sigma} \ln\frac{1+\sigma}{1-\sigma} \\
m_2(\sigma) &= \frac{1}{\sigma^2}(m_0(\sigma) - 1) \\
m_4(\sigma) &= \frac{1}{\sigma^4}\left(m_0(\sigma) - 1 - \frac{\sigma^2}{3}\right) \\
m_{00}(Q,s) &= m_0(\sigma) \\
m_{02}(Q,s) &= m_2(\sigma) + \frac{Q}{s} m_{11}(Q,s) \\
m_{04}(Q,s) &= m_4(\sigma) + \frac{Q^2}{\sigma^2}(5m_4(\sigma) - 3m_2(\sigma)) + m_{40}(Q,s) \\
m_{11}(Q,s) &= \frac{Qs}{2\sigma^2}(3m_2(\sigma) - m_0(\sigma)) \\
m_{13}(Q,s) &= \frac{Qs}{2\sigma^2}(5m_4(\sigma) - 3m_2(\sigma)) + m_{31}(Q,s) \\
m_{20}(Q,s) &= \frac{Q}{s} m_{11}(Q,s) \\
m_{22}(Q,s) &= \frac{Q^2}{4\sigma^2}(15m_4(\sigma) - 12m_2(\sigma) + m_0(\sigma)) + m_{40}(Q,s) \\
m_{31}(Q,s) &= \frac{Q^3 s}{8\sigma^4}(35m_4(\sigma) - 30m_2(\sigma) + 3m_0(\sigma)) \\
m_{40}(Q,s) &= \frac{Q}{s} m_{31}(Q,s)
\end{aligned} \quad (29)$$

We shall find it useful to further temporarily complicate matters by introducing yet another layer of indirection by use of the following linear combinations of the $m_n$ variables, as such:

$$\begin{aligned}
\phi_1(\sigma) &= 1 - \frac{3}{4}(m_0(\sigma) - m_2(\sigma)) \\
\phi_2(\sigma) &= -\frac{1}{\sigma^2}\left(1 - \frac{3}{2}(m_0(\sigma) - m_2(\sigma))\right) \\
\phi_3(\sigma) &= 1 - \frac{3}{4}(m_0(\sigma) + m_2(\sigma) - 2m_4(\sigma)) \\
\phi_4(\sigma) &= 1 - \frac{3}{8}(m_0(\sigma) + 2m_2(\sigma) + m_4(\sigma)) \\
\phi_5(\sigma) &= 1 - \frac{3}{2} m_2(\sigma)
\end{aligned} \quad (30)$$

these $\varphi_n$ functions will be called kernel functions.

### 3.1 Logarithmic Decomposition

Following Chandrasekhar [11] (e.g. p. 114 eqn 60), we will use so-called $H_n$ functions in order to solve these Laplacian kernel functions inside the integrals in equation (19). To achieve this goal, we define the $H_n$ functions in terms of the $\varphi_n$ functions:

$$\begin{aligned}
\phi_n(\sigma) &= \frac{1}{H_n(Q,s) H_n(Q,-s)} \\
\text{so} \log \phi_n(\sigma) &= -\log H_n(Q,s) - \log H_n(Q,-s)
\end{aligned} \quad (31)$$

This equation is also known as the Wiener-Hopf identity. $H_n(Q,s)$ is defined as analytic in the left half of the complex plane (i.e. for Re $s < 0$).

We shall require that the $\varphi_n$ kernel functions can be expressed as rational functions of the form:

$$\phi(\sigma) = \frac{\prod_{a=1}^{M}(\sigma^2 - z_a^2)}{\prod_{b=1}^{N}(\sigma^2 - p_b^2)} = \frac{\prod_{a=1}^{M}(\sigma - z_a)(\sigma + z_a)}{\prod_{b=1}^{N}(\sigma - p_b)(\sigma + p_b)} \quad (32)$$

where there are $2M$ zeros ($z_a$) and $2N$ poles ($p_b$) that are all in the right half plane (Re $z_a > 0$ and Re $p_b > 0$). This requirement allows us to find the $H_n(Q,s)$ function as:

$$H_n(Q,s) = \frac{\prod_{b=1}^{N}(s - \sqrt{(Q^2 + p_b^2)})}{\prod_{a=1}^{M}(s - \sqrt{(Q^2 + z_a^2)})} = \frac{H_{poles}}{H_{zeros}} \quad (33)$$

Once we have a potential solution $H_n(Q,s)$ in this form with poles in the numerator and zeros in the denominator, we can follow the Wiener-Hopf approach [13]. Taking the log of both sides we get

$$\log H_n(Q,s) = \log H_{poles} - \log H_{zeros} \quad (34)$$

Using Cauchy's integral formula, which is,

$$f(a) = \frac{1}{2\pi i} \oint \frac{f(z)}{z - s} dz \quad (35)$$

we can express (33) as the following complex contour integral using (31) and (34):

$$H_n(Q,s) = \exp\left(-\int \frac{dz}{2\pi i} \frac{\log \phi_n(z)}{z - s}\right) \quad (\text{Re } s < 0) \quad (36)$$

where the vertical contour is at Re $z = 0$. In addition, we carry out a transformation of $z = i \tan \beta$ to enable easier numerical evaluation:

$$H_s(Q,s) = \exp\left(\frac{s}{\pi} \int_0^{\pi/2} d\beta \frac{\ln \widetilde{\phi}_n(\beta)}{\sin^2 \beta + s^2 \cos^2 \beta}\right) \quad (\text{Re } s < 0) \quad (37)$$

with:

$$\widetilde{\phi}_1(\beta) = 1 - \frac{3}{4}((\cot^2\beta + 1)(\beta \cot\beta - 1) + 1)$$
$$\widetilde{\phi}_2(\beta) = \cot^2\beta\left(1 - \frac{3}{2}((\cot^2\beta + 1)(\beta \cot\beta - 1) + 1)\right)$$
$$\widetilde{\phi}_3(\beta) = 1 + \frac{3}{4}((2\cot^4\beta + \cot^2\beta - 1)(\beta \cot\beta - 1) + \frac{2}{3}\cot^2\beta - 1) \quad (38)$$
$$\widetilde{\phi}_4(\beta) = 1 - \frac{3}{8}((\cot^2\beta - 1)^2(\beta \cot\beta - 1) + \frac{1}{3}\cot^2\beta + 1)$$
$$\widetilde{\phi}_5(\beta) = 1 + \frac{3}{2}\cot^2\beta(\beta \cot\beta - 1)$$

and then we may make the replacement:

$$\beta \to \widetilde{\beta}(Q,\beta) = \arctan\sqrt{(Q^2 + \tan^2\beta)} \quad (39)$$

### 3.2 Derivation of bistatic matrix

**$\gamma_{44}$ element.** We are now faced with the task of deriving the bistatic matrix elements in (15) using the elements in equation (19) and (20). The easiest element to peel off is the $\gamma_{44}$ circular polarization element since it involves the fewest terms.

The equations for C and F in (20) constitute a 2x2 matrix that can be solved for $\gamma_{44}$. Using the Laplace form of these two parts of (20) and setting source terms to 1, we obtain:

$$C = \left(\frac{2}{3} - m_{02}\right)c - 2m_{11}f$$
$$F = \frac{3}{4}(m_{11}c + (m_{00} - m_{02} - m_{20})f) \quad (40)$$

Inserting the values for $m_{nn}$ using equation (29), the 2x2 matrix can be made diagonal by using these linear relationships, where the input parameters ($Q,s$) for c and f are made implicit:

$$\begin{bmatrix} \widetilde{c} \\ \widetilde{f} \end{bmatrix} = \begin{bmatrix} s & -2Q \\ Q/2 & -s \end{bmatrix} \begin{bmatrix} c \\ f \end{bmatrix} \quad (41)$$

and inverting this matrix for $c$ and $f$, and using equation (28) we obtain the following transfer matrix:

$$\begin{bmatrix} c \\ f \end{bmatrix} = \frac{1}{s^2 - Q^2} \begin{bmatrix} s & -2Q \\ Q/2 & -s \end{bmatrix} = \frac{1}{\sigma^2} \begin{bmatrix} s & -2Q \\ Q/2 & -s \end{bmatrix} \begin{bmatrix} \widetilde{c} \\ \widetilde{f} \end{bmatrix} \quad (42)$$

And the new functions obey:

$$\phi_5(\sigma)\widetilde{c}(Q,s) = \frac{-s}{(1-s)} I_4 + \widetilde{C}(Q,s)$$
$$\phi_1(\sigma)\widetilde{f}(Q,s) = \frac{Q}{(1-s)} I_4 + \widetilde{F}(Q,s) \quad (43)$$

where

$$\widetilde{C}(Q,s) = \int \frac{dt}{2\pi i(t-s)}(m_{02}c(Q,s) - 2m_{11}f(Q,s))$$
$$\widetilde{F}(Q,s) = \int \frac{dt}{2\pi i(t-s)} \frac{3}{4}(m_{11}c(Q,s) + (m_{00} - m_{02} - m_{20})f(Q,s) \quad (44)$$

Using the Wiener-Hopf procedure to solve equation and using equation (31) we obtain:

$$\widetilde{c}(Q,s) = \left(c_1 + \frac{c_2}{1-s}\right) I_4 H_5(Q,s)$$
$$\widetilde{f}(Q,s) = \left(f_1 + \frac{f_2}{1-s}\right) I_4 H_1(Q,s) \quad (45)$$

We must now find the Wiener-Hopf constants $c_1$, $c_2$, $f_1$ and $f_2$. Taking the $s \to 1$ limit for equation (42) constrains two of them:

$$c_2 = \frac{3}{2} I_4 H_5(Q,-1)$$
$$f_2 = \frac{3}{4} Q I_4 H_1(Q,s) \quad (46)$$

$c_1$, and $f_1$ are then fixed by letting $\sigma \to 0$ (i.e. $s \to Q$ and $s \to -Q$) and requiring that c(Q,s) and f(Q,s) in equation (42) remain finite. After some significant manipulations, we arrive at:

$$\gamma_{44}(Q) = \frac{-c(Q,-1)}{I_4} = -\frac{3}{4} \frac{N_{44}(Q)}{(1-Q^2)^2 (H_1^2(Q,-Q) + H_5^2(Q,-Q))}$$

where

$$\begin{aligned}
N_{44}(Q) = &\, Q^2(1+Q)^2 H_5^2(Q,-Q) H_1^2(Q,-1) + \\
&\, Q^2(1-Q)^2 H_1^2(Q,-Q) H_1^2(Q,-1) + \\
&\, (1-Q)^2 H_5^2(Q,-Q) H_5^2(Q,-1) + \\
&\, (1+Q)^2 H_1^2(Q,-Q) H_5^2(Q,-1) \\
&\, -8Q^2 H_1(Q,-Q) H_5(Q,-Q) H_1(Q,-1) H_5(Q,-1)
\end{aligned} \quad (47)$$

**$\gamma_{33}$ element.** We now turn to the next easiest bistatic matrix element of equation (15) which is $\gamma_{44}$. As for the derivation of $\gamma_{44}$, we extract a 2x2 matrix in (20) for E and H or their Laplace transform counterparts $e(Q,s)$ and $h(Q,s)$. To solve the 2x2 matrix, we use the following linear combinations of $e(Q,s)$ and $h(Q,s)$, where the input parameters ($Q,s$) for e and h are made implicit:

$$\begin{aligned}
\tilde{e} &= -se + 2Qh \\
\tilde{h} &= \frac{-Q}{2} e + sh
\end{aligned} \quad (48)$$

And in a similar manner as for equation (43) the new functions obey:

$$\begin{aligned}
\phi_3(\sigma) \tilde{e}(Q,s) &= -\frac{Q}{1-s} I_3 + \tilde{E}(Q,s) \\
\phi_4(\sigma) \tilde{h}(Q,s) &= -\frac{s}{1-s} I_3 + \tilde{H}(Q,s)
\end{aligned} \quad (49)$$

The solution of this equation is analogous to that given in equations (44) to (47). We eventually arrive at:

$$\gamma_{33}(Q) = -\frac{3}{4} \frac{N_{33}(Q)}{(1-Q^2)^2 (H_3^2(Q,-Q) + H_4^2(Q,-Q))}$$

where

$$\begin{aligned}
N_{33}(Q) = &\, Q^2(1+Q)^2 H_4^2(Q,-Q) H_3^2(Q,-1) + \\
&\, Q^2(1-Q)^2 H_3^2(Q,-Q) H_3^2(Q,-1) + \\
&\, (1-Q)^2 H_4^2(Q,-Q) H_4^2(Q,-1) + \\
&\, (1+Q)^2 H_1^2(Q,-Q) H_5^2(Q,-1) \\
&\, -8Q^2 H_3(Q,-Q) H_4(Q,-Q) H_3(Q,-1) H_4(Q,-1)
\end{aligned} \quad (50)$$

**$\gamma_{11}, \gamma_{12}, \gamma_{22}$ elements.** The most difficult bistatic matrix elements are now solved. The 4 dependent relations of equation (20) for A, B, D and G form a 4x4 matrix system that must be solved. This presents a challenging booking problem. As a first step, the following linear combinations of the Laplace form of A, B, D and G are put in the following linear combinations, where the input parameters ($Q,s$) for a, b, d, g are implicit:

temporarily $P = -2\sigma^4 - 2Q^2\sigma^2 + 2\sigma^2 + 3Q^2$

$$\begin{bmatrix} \tilde{a} \\ \tilde{b} \\ \tilde{d} \\ \tilde{g} \end{bmatrix} = \begin{bmatrix} 2\sigma^2 & -Q^2 & -2Qs & 2Q^2 \\ -2\sigma^4 & P & 2Qs(-2\sigma^2+3) & 2Q^2(2\sigma^2-3) \\ 0 & Qs & \sigma^2+2Q^2 & -2Qs \\ 0 & Q^2/2 & Qs & -(2\sigma^2+Q^2) \end{bmatrix} \begin{bmatrix} a \\ b \\ d \\ g \end{bmatrix} \quad (51)$$

And in a similar manner to equations (43) and (49), the new functions obey:

$$\begin{aligned}
\phi_1(\sigma) \tilde{a}(Q,s) &= \frac{2(s^2-1)(I_1+I_2) - 2Q^2 I_2}{1-s} + \tilde{A}(Q,s) \\
\phi_2(\sigma) \tilde{b}(Q,s) &= \frac{2Q^2 I_1}{1-s} + \tilde{B}(Q,s) \\
\phi_3(\sigma) \tilde{d}(Q,s) &= \frac{2Qs I_1}{1-s} + \tilde{D}(Q,s) \\
\phi_4(\sigma) \tilde{g}(Q,s) &= \frac{2s^2(I_1-I_2) + 2Q^2 I_2}{1-s} + \tilde{G}(Q,s)
\end{aligned} \quad (52)$$

Inverting equation (51) with respect to a, b, d and g and using equation (31), where again the input parameters ($Q,s$) for a, b, d, g and their tilde counterparts are implicit, we get the following transfer matrix:

temporarily, $P = 4\sigma^2(\sigma^2-1)$, $S = \sigma^2-1$ and $R = 2\sigma^2 + Q^2$

$$\begin{bmatrix} Pa \\ Pb \\ \frac{P}{2}d \\ 2Pg \end{bmatrix} = \begin{bmatrix} \sigma^2(R-2) & Q^2 & 4QsS & -2Q^2 S \\ -\sigma^2(R+Q^2) & -(R+Q^2) & -12QsS & 6Q^2 S \\ Qs\sigma^2 & Qs & 2(\sigma^2+2Q^2)S & -2QsS \\ Q^2\sigma^2 & Q^2 & 4QsS & -2RS \end{bmatrix} \begin{bmatrix} \tilde{a} \\ \tilde{b} \\ \tilde{d} \\ \tilde{g} \end{bmatrix} \quad (53)$$

So that we now have, setting $s = -1$ for the backscattering direction:

temporarily $P = -(1-Q^2)^{-2}$

$$\begin{bmatrix} \gamma_{11}(Q) & \gamma_{12}(Q) \\ \gamma_{21}(Q) & \gamma_{22}(Q) \end{bmatrix} \begin{bmatrix} I_1 \\ I_2 \end{bmatrix} = \begin{bmatrix} 0 & \frac{P}{2} & 2QP & P \\ \frac{-P^{1/2}}{2} & 0 & 0 & -P^{1/2} \end{bmatrix} \begin{bmatrix} \tilde{a}(Q,-1) \\ \tilde{b}(Q,-1) \\ \tilde{d}(Q,-1) \\ \tilde{g}(Q,-1) \end{bmatrix} \quad (54)$$

The solution to equation equation (52) is:

$$\begin{aligned}
\tilde{a}(Q,s) &= \left(a_1 + a_2 s + \frac{a_3}{1-s}\right) H_1(q,s) \\
\tilde{b}(Q,s) &= \left(b_1 + b_2 s + b_3 s^2 + \frac{b_4}{1-s}\right) H_2(q,s) \\
\tilde{d}(Q,s) &= \left(d_1 + d_2 s + \frac{d_3}{1-s}\right) H_3(q,s) \\
\tilde{g}(Q,s) &= \left(g_1 + g_2 s + \frac{g_3}{1-s}\right) H_4(q,s)
\end{aligned} \quad (55)$$

Where $a_1$ to $g_3$ are constants to be found using the Wiener-Hopf approach.

Taking the $s \to 1$ limit for equation (52) constrains four of them:

$$\begin{aligned}
a_3 &= \frac{-3}{2} Q^2 I_2 H_1(Q,-1) \\
b_4 &= -3Q^2 I_1 H_2(Q,-1) \\
d_3 &= \frac{3}{2} Q I_1 H_3(Q,-1) \\
g_3 &= \frac{-3}{4} (I_1 + (Q^2-1) I_2) H_4(Q,-1)
\end{aligned} \quad (56)$$

The last nine functional coefficients are constrained by assuming that the functions $a(Q,s)$, $b(Q,s)$, $d(Q,s)$, $g(Q,s)$ have the expected regularity properties at the points where the inversion formulas (equation (53)) are singular, that is when $P = 0$, which occurs when 1.), $\sigma^2=0$, or using equation (31), $s=+/-Q$ or 2.) when $\sigma^2-1=0$ or using equation (31), $s=+/-\sqrt{(1+Q^2)}$. We thus obtain:

$$\begin{aligned}
\tilde{a}(Q,-\sqrt{(1+Q^2)})+\tilde{b}(Q,-\sqrt{(1+Q^2)})&=0 \\
\tilde{a}(Q,\sqrt{(1+Q^2)})+\tilde{b}(Q,\sqrt{(1+Q^2)})&=0 \\
\tilde{a}(Q,-Q)+2\tilde{g}(Q,-Q)&=0 \\
\tilde{b}(Q,-Q)-6\tilde{g}(Q,-Q)&=0 \\
\tilde{d}(Q,-Q)+2\tilde{g}(Q,-Q)&=0 \\
\frac{\partial}{\partial s}(\tilde{b}(Q,s)+4\tilde{d}(Q,s)+2\tilde{g}(Q,s))_{s=-Q}-8Q\tilde{g}(Q,-Q)&=0 \quad (57)\\
\tilde{b}(Q,Q)-6\tilde{g}(Q,Q)&=0 \\
\tilde{d}(Q,Q)-2\tilde{g}(Q,Q)&=0 \\
3\frac{\partial}{\partial s}(\tilde{b}(Q,s)-4\tilde{d}(Q,s)+2\tilde{g}(Q,s))_{s=Q}+10Q\tilde{a}(Q,Q)+44Q\tilde{g}(Q,Q)&=0
\end{aligned}$$

By inserting the elements of equation (55) and (56) in the relations of equation (57) we can theoretically find equations relating the Wiener-Hopf coefficients $a_1, a_2, a_3$, $b_1, b_2, b_3, b_4$, $d_1, d_2, d_3$, $g_1, g_2$ and $g_3$. In fact, we are left with a 9x9 linear matrix problem which must be solved with symbolic manipulation software, such as Maxima (maxima.sourceforge.net). This is the approach taken in this paper in order to derive the results discussed in the next section.

## 4. Results

*4.1 Linear Polarization Comparison*

If we wish to express the results of Section 1 in terms of a different set of Stokes Parameters, for example, the 'standard' Stokes Parameters defined as:

$$\begin{aligned}
I&=\langle E_\parallel E_\parallel^*\rangle+\langle E_\perp E_\perp^*\rangle \\
Q&=\langle E_\parallel E_\parallel^*\rangle-\langle E_\perp E_\perp^*\rangle \\
U&=\langle E_\parallel E_\perp^*\rangle+\langle E_\perp E_\parallel^*\rangle=2\,\mathrm{Re}\langle E_\theta E_\phi^*\rangle \\
V&=i(\langle E_\parallel E_\perp^*\rangle+\langle E_\perp E_\parallel^*\rangle)=2\,\mathrm{Im}\langle E_\theta E_\phi^*\rangle
\end{aligned} \quad (58)$$

It is necessary to use a transfer matrix to convert $I_1$ and $Q_1$ of equation (1) into $I$ and $Q$ of (58). Effectively we are modifying equations (16), (17) and (18) in terms of the 'standard' Stokes Parameters. This has been done previously by Mishchenko et al. [2] who arrived at the following result:

$$S_{nn}(\theta)=S^1(\theta)+S^L(\theta)+S^C(\theta)$$

where

$$S^1(\alpha)\approx S^1(0)=\frac{3}{16}\begin{bmatrix} 1 & 0 & 0 & 0 \\ 0 & 1 & 0 & 0 \\ 0 & 0 & -1 & 0 \\ 0 & 0 & 0 & -1 \end{bmatrix}$$

$$S^L(\alpha)\approx S^L(0)=\frac{1}{4}\begin{bmatrix} \gamma_{11}(0)+\gamma_{22}(0) & 0 & 0 & 0 \\ 0 & \gamma_{11}(0)-\gamma_{12}(0) & 0 & 0 \\ 0 & 0 & \gamma_{12}(0)-\gamma_{11}(0) & 0 \\ 0 & 0 & 0 & \gamma_{44}(0) \end{bmatrix}-S^1$$

$$S^C(\alpha)\approx S^C(Q)=\begin{bmatrix} S^C_{11}(Q) & S^C_{12}(Q) & 0 & 0 \\ S^C_{12}(Q) & S^C_{22}(Q) & 0 & 0 \\ 0 & 0 & S^C_{33}(Q) & 0 \\ 0 & 0 & 0 & S^C_{44}(Q) \end{bmatrix}-S^1(0) \quad (59)$$

where

$$S^C_{11}(Q)=\frac{1}{8}[\gamma_{11}(Q)+\gamma_{22}(Q)-\gamma_{33}(Q)+\gamma_{44}(Q)]$$
$$S^C_{22}(Q)=\frac{1}{8}[\gamma_{11}(Q)+\gamma_{22}(Q)+\gamma_{33}(Q)-\gamma_{44}(Q)]$$
$$S^C_{33}(Q)=\frac{1}{8}[\gamma_{33}(Q)+\gamma_{44}(Q)]-\frac{1}{4}\gamma_{12}(Q)$$
$$S^C_{44}(Q)=\frac{1}{8}[\gamma_{33}(Q)+\gamma_{44}(Q)]+\frac{1}{4}\gamma_{12}(Q)$$
$$S^C_{12}(Q)=\frac{1}{8}[\gamma_{11}(Q)-\gamma_{22}(Q)]$$

The enhancement factor and degree of linear polarization are given by the following equations:

$$\begin{aligned}
\zeta(\alpha)&=\frac{S^1_{11}(0)+S^L_{11}(0)+S^C_{11}(0)}{S^1_{11}(0)+S^L_{11}(0)} \\
DLOP(\alpha)&=\frac{-Q(\alpha)}{I(\alpha)}=\frac{-S^C_{12}(Q)}{S^1_{11}(0)+S^L_{11}(0)+S^C_{11}(Q)}
\end{aligned} \quad (60)$$

We first calculated the enhancement factor $\zeta$ and degree of linear polarization (DLOP) results of equation (60) in order to test our code was functioning correctly. Our results are shown in Figure 2. The angular parameter Q is given by equation (11).

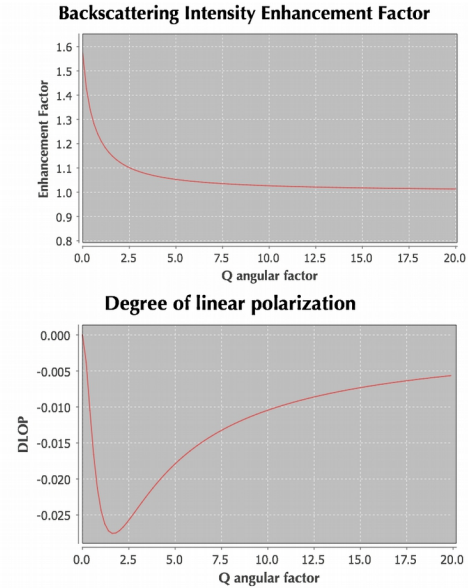

Fig. 2. Comparison of Mishchenko et al. [2] linear polarization results for enhancement factor and degree of linear polarization.

*4.2 New Circular Polarization Results*

Amic et al. [1] considered four enhancement factors for cases of input linearly polarized light with parallel and orthogonal channels. They considered the use of circularly polarized beams by plotting the helicity preserved and opposite helicity enhancement factors.

Here we use the commonly adopted circular polarization ratio using the standard Stokes parameters given in (58), for easy comparison with laboratory or planetary results [14].

In order to calculate the circular polarization ratio we use the follow relation:

$$\mu_c(Q) = \frac{S_{11}(Q) + S_{44}(Q)}{S_{11}(Q) - S_{44}(Q)} \quad (61)$$

The plot of the circular polarization is given in Fig.3.

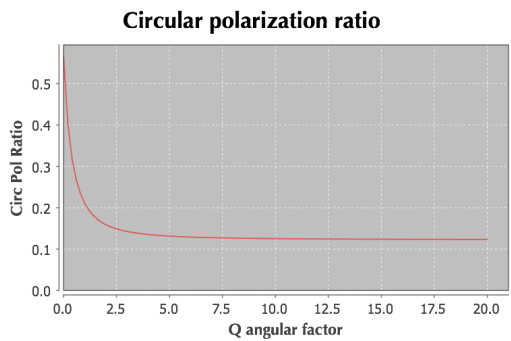

Fig. 3. Circular Polarization Ratio enhancement factor versus angular factor.

Brown and Xie [15] outlined an approach of using a stereographic projection onto a flat plane in order to map the Müller matrix scattering maps and using inherent symmetries of scatterers to check on the results. We have plotted the Müller matrix hemispherical map for the Cooperon contribution ($S^C$) in equation (59) to the scattered light in Fig.4.

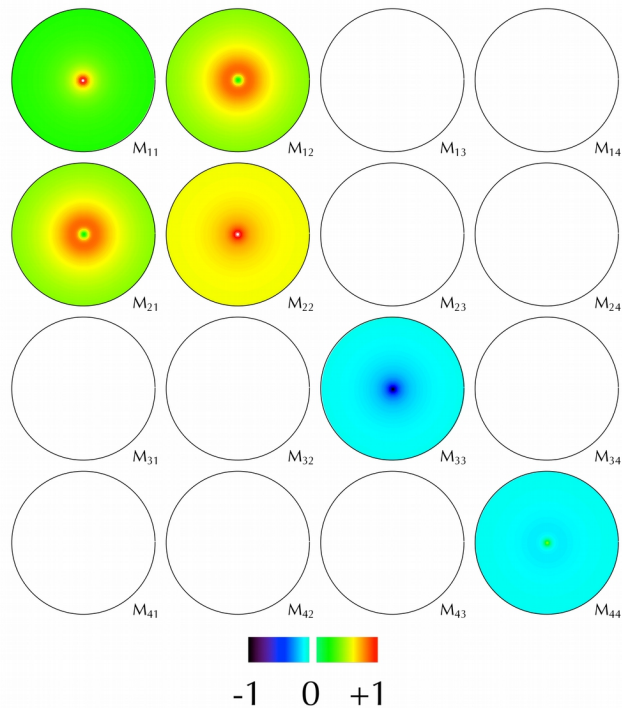

Fig. 4. Hemispherical map for the cyclical (Cooperon) contribution ($S^C$) to the scattered light versus angular factor (stretched from q=0 in center to q=10 at rim) for Rayleigh scattering. The radial symmetry of this pattern reflects the symmetry of the Rayleigh scattering component in this situation. Note that the white regions indicate the value zero. Elements $M_{13}$, $M_{14}$, $M_{23}$, $M_{24}$, $M_{31}$, $M_{32}$, $M_{41}$, $M_{42}$, $M_{43}$ and $M_{34}$ are all zero due to symmetries of the model setup.

*Cooperon Müller Matrix*. The $M_{11}$ element of Figure 4 displays a sharp positive peak at backscatter, corresponding to the well known intensity coherent backscattering effect. The $M_{12}$ and $M_{21}$ elements are identical and display a positive slope in a peak shape from the rim toward the center, which is terminated by a rolloff and negative slope at backscatter. The $M_{33}$ element displays a negative peak, less sharp than the $M_{11}$ peak. The $M_{44}$ element displays a negative peak from the rim, which is again terminated by a weak positive slope at close to backscatter (q=0).

*'Planetary' geometry*. In order to further close the gap between theory and experiment, and make our results more useful for comparison to laboratory work, we use equation (9) of Brown [14] to to derive the expected backscattering patterns for the 'planetary' observational geometry. We plug equation (59) into equation (9) of Brown [14] in order to calculate the backscattering Müller matrix.

As can be seen from Figure 5, the strong symmetrical patterns of the backscattering LIDAR as still very prominently observed in the laboratory/LIDAR geometry. These observational effects have to be disentangled in any future observational program.

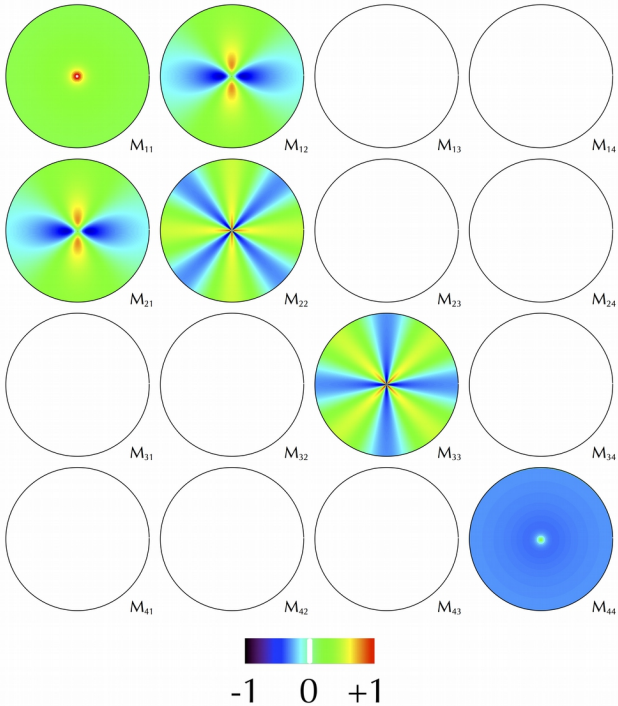

Fig. 5. Cooperon contribution for Rayleigh scattering in 'planetary' scattering geometry as described in [14]. Note that the white regions indicate the value zero. Elements $M_{13}$, $M_{14}$, $M_{23}$, $M_{24}$, $M_{31}$, $M_{32}$, $M_{41}$, $M_{42}$, $M_{43}$ and $M_{34}$ are all zero due to symmetries of the model setup.

## 5. Discussion

### 5.1 Comparison of backscattered intensity, linear polarization and circular polarization peaks

It is important to emphasize the key result of this paper – we have isolated a third backscattering effect to be investigated and exploited by future laboratory and fields experiments.

The circular polarization backscattering peak shown in Figure 3 can be compared with the intensity and linear polarization negative backscattering features in Figure 2. For the non absorbing Rayleigh scattering semi infinite slab of particles case investigated here, we find the circular polarization peak is a little narrower that the intensity peak, and also a little weaker. Both of these features are stronger in intensity than the negative linear polarization peak, however the negative linear polarization peak is far wider than either the intensity or circular polarization. Of course these features will change shape when larger particles are observed, and also as asymmetric scattering, internal scattering, and absorbing particles are investigated [16]. Future experimental and theoretical studies should address whether the relative intensities and peak widths remain as predicted in this paper.

### 5.2 Comparison to Monte Carlo results

Muinonen [17] used a Monte Carlo technique to compute the coherent backscattering of isolated sphere targets and provided a numerical solution for the circular polarization peak that we have calculated analytically here. His Fig. 9(c) matches our Fig. 3 relatively well. In addition, Muinonen was able to show the variation of the peak with size and vary the absorption factor. Muinonen and Videen [18] recently employed a Monte Carlo code to investigate the backscattering behavior of a slab composed of clusters of non absorbing ($m$=1.5) Mie spheres. Figure 2 of their paper shows the circular polarization ratio as a function of backscattering angle for a range of sphere sizes and two packing densities (3% and 6%). Their circular polarization peaks show significant structure as the grain size is increased. The peak width is qualitatively consistent with our result because it is narrower than the intensity peak, which are both narrower than the linear polarization peak.

Kuz'min and Meglinski [19] have carried out work on a Monte Carlo based approach to modeling the backscattering enhancement factor due to Cooperons, providing an examination of the circular co-polarized and cross-polarized backscattered light. This differs from our work, where we have examined the full Müller matrix response.

### 5.3 Comparison to other Schwarzchild-Milne based approaches

Kuz'min and Askenova [20] developed a Schwarzchild-Milne equation based approach which inserted a first degree Legendre polynomial ($P_1$) asymmetric phase function and examined the effect this had on linear depolarization of backscattered light. This approach of using a spherical orthogonal representation to deal with strongly asymmetric scattering was also applied by Amic et al. [21].

### 5.4 Applications in Astronomy

Recent hardware advances in the astronomical world have made it possible to obtain linear polarization of debris disks and protoplanetary disks and potentially observe the negative peak of polarization related to coherent backscattering of these bodies [22]. This makes it even more important to understand the linkages between CB and polarization peaks. This study has also shown the importance of the circular polarization peak and motivates further work on circular polarization measurement strategies [23].

## 6. Conclusions

### 6.1 Implications for Experimental Work

The results uncovered here will be of interest to other workers attempting to understand the circular polarization of Rayleigh scatterers molecules and solid particles [24] by using the coherent backscattering of light. Applications of this work to hyperspectral [25], [26] and hyperspectral polarization sensitive datasets [27], [28] will also be of interest, and measurements of the Cooperon contribution in Figures 4 and 5 have the potential to be tested in future experimental endeavors.

## 6.2 Future Work

In this paper, in keeping with previous studies, we have presented a theoretical model of the 'planetary' geometric scattering situation. Now that the 'laboratory/LIDAR' scattering geometry has been recognized [14], one is tempted to think the new scattering geometry might potentially be derived using equation (13) of Brown [14], however, in fact this will also require re-working of the Chandrasekhar-inspired phase functions which were developed in the 'planetary' geometry by him. These are shown here as equation (8). We therefore save this application for future work.

## Acknowledgements

This work was sponsored in part by NASA Grant Planetary Geology and Geophysics NNX 11AP23G administered by Discipline Scientist Michael S. Kelley. I would like to thank Tanya Christensen for her invaluable assistance in working on and checking the analysis in manuscript, and Jean-Marc Luck for providing a Fortran code for inversion of equation (57).

## Appendix A

We give here a partial derivation of equations (19) and (20). We only derive the $\Gamma^{(0)}$ (k=0) term of (19) and the A term of equation (20) as an example of how the other parameters in those equations may be derived.

We start with equation (14) which has a non-integral term and two relevant integral terms. To deal with the non-integral term, we first set k=0 in the first term in equation (14) and insert the $P^{(0)}$ term from equation (8) to get:

$$P^{(0)}(\mu,\mu_a) \cdot I_a e^{-i(0)\varphi_a - \tau/\mu_a} = P^{(0)}(\mu,\mu_a) \cdot I_a e^{-\tau/\mu_a}$$

$$= \frac{3}{4} \begin{pmatrix} 2(1-\mu^2)(1-\mu_a^2)+\mu^2\mu_a^2 & \mu^2 & 0 & 0 \\ \mu_a^2 & 1 & 0 & 0 \\ 0 & 0 & 0 & 0 \\ 0 & 0 & 0 & 2\mu\mu_a \end{pmatrix} \begin{pmatrix} I_1 \\ I_2 \\ I_3 \\ I_4 \end{pmatrix} e^{-\tau/\mu_a}$$

$$= \frac{3}{4} \begin{pmatrix} [2(1-\mu^2)(1-\mu_a^2)+\mu^2\mu_a^2]I_1 + [\mu^2]I_2 \\ \mu_a^2 I_1 + I_2 \\ 0 \\ 2\mu\mu_a I_4 \end{pmatrix} e^{-\tau/\mu_a} \quad (62)$$

The first integral term in equation (14) is:

$$\int_0^\tau \int_0^1 \frac{e^{-(\tau-\tau')/\mu'}}{2\mu'} P^{(k)}(\mu,\mu') \sum_{j=-2}^{2} i^{k-j} J_{k-j}(Q(\tau-\tau')\nu'/\mu') \Gamma^{(j)}(\tau',\mu') \partial\mu' \partial\tau' \quad (63)$$

We first evaluate the $P^{(k)}J_{k-j}\Gamma^{(k)}$ kernel. Set k = 0 and use the $P^{(0)}$ term from equation (8) and the properties of Bessel functions in equation (13):

$$P^{(0)}(\mu,\mu') \sum_{j=-2}^{2} i^{-j} J_{-j}(Q(\tau-\tau')\nu'/\mu') \Gamma^{(j)}(\tau',\mu')$$

$$= \frac{3}{4} \begin{pmatrix} 2(1-\mu^2)(1-\mu'^2)+\mu^2\mu'^2 & \mu^2 & 0 & 0 \\ \mu'^2 & 1 & 0 & 0 \\ 0 & 0 & 0 & 0 \\ 0 & 0 & 0 & 2\mu\mu' \end{pmatrix} \begin{pmatrix} \sum_{j=-2}^{2} i^{-j} J_{-j}(z) \Gamma_1^{(j)}(\tau',\mu') \\ \sum_{j=-2}^{2} i^{-j} J_{-j}(z) \Gamma_2^{(j)}(\tau',\mu') \\ \sum_{j=-2}^{2} i^{-j} J_{-j}(z) \Gamma_3^{(j)}(\tau',\mu') \\ \sum_{j=-2}^{2} i^{-j} J_{-j}(z) \Gamma_4^{(j)}(\tau',\mu') \end{pmatrix}$$

$$= \frac{3}{4} \begin{pmatrix} 2(1-\mu^2)(1-\mu'^2)+\mu^2\mu'^2 & \mu^2 & 0 & 0 \\ \mu'^2 & 1 & 0 & 0 \\ 0 & 0 & 0 & 0 \\ 0 & 0 & 0 & 2\mu\mu' \end{pmatrix} \begin{pmatrix} i^2 J_2 \Gamma_1^{(-2)} + i^1 J_1 \Gamma_1^{(-1)} + i^0 J_0 \Gamma_1^{(0)} + i^{-1} J_{-1} \Gamma_1^{(1)} + i \\ i^2 J_2 \Gamma_2^{(-2)} + i^1 J_1 \Gamma_2^{(-1)} + i^0 J_0 \Gamma_2^{(0)} + i^{-1} J_{-1} \Gamma_2^{(1)} + i \\ i^2 J_2 \Gamma_3^{(-2)} + i^1 J_1 \Gamma_3^{(-1)} + i^0 J_0 \Gamma_3^{(0)} + i^{-1} J_{-1} \Gamma_3^{(1)} + i \\ i^2 J_2 \Gamma_4^{(-2)} + i^1 J_1 \Gamma_4^{(-1)} + i^0 J_0 \Gamma_4^{(0)} + i^{-1} J_{-1} \Gamma_4^{(1)} + i \end{pmatrix} \quad (64)$$

Noting that $i^0 = 1$ and from (4) $i^{-2}J_{-2} = i^2 J_2$ and $i^{-1}J_{-1} = -i^1 J_{-1} = i^1 J_1$ this simplifies significantly. We use the following shorthand:

$$\begin{aligned} \Gamma_1 &= -J_2(z)(\Gamma_1^{(-2)}+\Gamma_1^{(2)}) + iJ_1(z)(\Gamma_1^{(-1)}+\Gamma_1^{(1)}) + J_0(z)\Gamma_1^{(0)} \\ \Gamma_2 &= -J_2(z)(\Gamma_2^{(-2)}+\Gamma_2^{(2)}) + iJ_1(z)(\Gamma_2^{(-1)}+\Gamma_2^{(1)}) + J_0(z)\Gamma_2^{(0)} \\ \Gamma_3 &= -J_2(z)(\Gamma_3^{(-2)}+\Gamma_3^{(2)}) + iJ_1(z)(\Gamma_3^{(-1)}+\Gamma_3^{(1)}) + J_0(z)\Gamma_3^{(0)} \\ \Gamma_4 &= -J_2(z)(\Gamma_4^{(-2)}+\Gamma_4^{(2)}) + iJ_1(z)(\Gamma_4^{(-1)}+\Gamma_4^{(1)}) + J_0(z)\Gamma_4^{(0)} \end{aligned} \quad (65)$$

Then we get:

$$P^{(0)}(\mu,\mu')\Gamma^{(0)}(\tau',\mu') = \frac{3}{4} \begin{pmatrix} 2(1-\mu^2)(1-\mu'^2)+\mu^2\mu'^2 & \mu^2 & 0 & 0 \\ \mu'^2 & 1 & 0 & 0 \\ 0 & 0 & 0 & 0 \\ 0 & 0 & 0 & 2\mu\mu' \end{pmatrix} \begin{pmatrix} \Gamma_1 \\ \Gamma_2 \\ \Gamma_3 \\ \Gamma_4 \end{pmatrix}$$

$$= \frac{3}{4} \begin{pmatrix} (2(1-\mu^2)(1-\mu'^2)+\mu^2\mu'^2)\Gamma_1 + \mu^2 \Gamma_2 \\ \mu'^2 \Gamma_1 + \Gamma_2 \\ 0 \\ 2\mu\mu' \Gamma_4 \end{pmatrix}$$

$$= \frac{3}{4} \begin{pmatrix} (\mu^2\mu'^2)\Gamma_1 + \mu^2 \Gamma_2 + ((1-\mu^2)2(1-\mu'^2))\Gamma_1 \\ \mu'^2 \Gamma_1 + \Gamma_2 \\ 0 \\ 2\mu\mu' \Gamma_4 \end{pmatrix} \quad (66)$$

Let A' = $\mu'^2 \Gamma_1 + \Gamma_2$ and B' = $2(1-\mu'^2)\Gamma_1$ and C' = $2\mu'\Gamma_4$ to get:

$$P^{(0)}(\mu,\mu')\Gamma^{(0)}(\tau',\mu') = = \frac{3}{4} \begin{pmatrix} \mu^2 A' + (1-\mu^2)B' \\ A' \\ 0 \\ \mu C' \end{pmatrix} \quad (67)$$

We then substitute (67), (63) and (62) into equation (14) for k=0 and we eventually get:

$$\begin{pmatrix} \Gamma_1^{(0)} \\ \Gamma_2^{(0)} \\ \Gamma_3^{(0)} \\ \Gamma_4^{(0)} \end{pmatrix} = \frac{3}{4} \begin{pmatrix} [2(1-\mu^2)(1-\mu_a^2)+\mu^2\mu_a^2]I_1 + [\mu^2]I_2 \\ \mu_a^2 I_1 + I_2 \\ 0 \\ 2\mu\mu_a I_4 \end{pmatrix} e^{-\tau/\mu_a} + \frac{3}{4} \int_0^\infty \int_0^1 \frac{e^{-|\tau-\tau'|/\mu'}}{2\mu'} \begin{pmatrix} \mu^2 A' + (1-\mu^2)B' \\ A' \\ 0 \\ \mu C' \end{pmatrix} \partial \quad (68)$$

Now let's concentrate on the top line for $\Gamma_1^{(0)}$ and collect terms of $1-\mu^2$ and $\mu^2$:

$$\begin{aligned}\Gamma_1^{(0)}\frac{4}{3}&=[2(1-\mu^2)(1-\mu_a^2)+\mu^2\mu_a^2]I_1+[\mu^2]I_2]e^{-\tau/\mu_a}+\int_0^\infty\int_0^1\frac{e^{-|\tau-\tau'|/\mu'}}{2\mu'}(\mu^2A'+(1-\mu^2)B')\partial\mu'\partial\tau\\&=(1-\mu^2)[2I_1(1-\mu_a^2)e^{-\tau/\mu_a}+\int_0^\infty\int_0^1\frac{e^{-|\tau-\tau'|/\mu'}}{2\mu'}B'\partial\mu'\partial\tau']\\&+(\mu^2)[(\mu_a^2I_1+I_2)e^{-\tau/\mu_a}+\int_0^\infty\int_0^1\frac{e^{-|\tau-\tau'|/\mu'}}{2\mu'}A'\partial\mu'\partial\tau']\end{aligned} \quad (69)$$

Note that in the last line, the multiplier of $\mu^2$ is equal to the second line of (68), which is $\Gamma_2^{(0)}$ – so we set this equal to A, $\Gamma_2^{(0)}=A$. If we set the multiplier of $(1-\mu^2)$ in the last line of (69) to $\widetilde{B}$, we get the relation:

$$\begin{aligned}\Gamma_1^{(0)}&=(1-\mu^2)\widetilde{B}+(\mu^2)A\\&=(1-\mu^2)\widetilde{B}+(\mu^2-1+1)A\\&=(1-\mu^2)(\widetilde{B}-A)+A\\&=(1-\mu^2)B+A\end{aligned} \quad (70)$$

So we therefore have:

$$\begin{aligned}\Gamma_1^{(0)}(\tau,\mu)&=A(\tau)+B(\tau)(1-\mu^2)\\\Gamma_2^{(0)}(\tau,\mu)&=A(\tau)\end{aligned} \quad (71)$$

To evaluate $A(\tau)$, we set it equal to line 2 of (68):

$$\Gamma_2^{(0)}(\tau,\mu)=A(\tau)=\frac{3}{4}[(\mu_a^2I_1+I_2)e^{-\tau/\mu_a}+\int_0^\infty\int_0^1\frac{e^{-|\tau-\tau'|/\mu'}}{2\mu'}(\mu'^2\Gamma_1+\Gamma_2)\partial\mu'\partial\tau'] \quad (72)$$

Substitute the top two lines of (68) into (72):

$$\begin{aligned}\Gamma_2^{(0)}(\tau,\mu)=A(\tau)&=\frac{3}{4}(\mu_a^2I_1+I_2)e^{-\tau/\mu_a}\\&+\frac{3}{4}\int_0^\infty\int_0^1\frac{e^{-|\tau-\tau'|/\mu'}}{2\mu'}\mu'^2[-J_2(z)(\Gamma_1^{(-2)}+\Gamma_1^{(2)})+iJ_1(z)(\Gamma_1^{(-1)}+\Gamma_1^{(1)})+J_0(z)\Gamma_1^{(0)}]\partial\mu'\partial\tau'\\&+\frac{3}{4}\int_0^\infty\int_0^1\frac{e^{-|\tau-\tau'|/\mu'}}{2\mu'}[-J_2(z)(\Gamma_2^{(-2)}+\Gamma_2^{(2)})+iJ_1(z)(\Gamma_2^{(-1)}+\Gamma_2^{(1)})+J_0(z)\Gamma_2^{(0)}]\partial\mu'\partial\tau'\end{aligned} \quad (73)$$

Using (71) we know $\Gamma_1^{(0)}=A(\tau)+B(\tau)(1-\mu^2)$ and $\Gamma_2^{(0)}=A(\tau)$.

Invoking knowledge outside this Appendix, (but for a proof, the reader need only carry out this analysis for k=1 and 2), we have:

$$\begin{aligned}\Gamma_1^{(1)}(\tau,\mu)&=-\Gamma_1^{(-1)*}(\tau,\mu)=\mu v(iD(\tau)-E(\tau))\\\Gamma_2^{(1)}(\tau,\mu)&=-\Gamma_2^{(-1)*}(\tau,\mu)=0\\\Gamma_1^{(2)}(\tau,\mu)&=\Gamma_1^{(-2)*}(\tau,\mu)=\mu^2(G(\tau)+iH(\tau))\\\Gamma_2^{(2)}(\tau,\mu)&=\Gamma_2^{(-2)*}(\tau,\mu)=-(G(\tau)+iH(\tau))\end{aligned} \quad (74)$$

so, since (*) is the complex conjugate operator, this implies:

$$\begin{aligned}-\Gamma_1^{(-1)}(\tau,\mu)&=\mu v(-iD(\tau)-E(\tau))=-\mu v(iD(\tau)+E(\tau))\\-\Gamma_2^{(-1)}(\tau,\mu)&=0\\\Gamma_1^{(-2)}(\tau,\mu)&=\mu^2(G(\tau)-iH(\tau))\\\Gamma_2^{(-2)}(\tau,\mu)&=-(G(\tau)-iH(\tau))\end{aligned} \quad (75)$$

Substituting these relations in, we get:

$$\begin{aligned}A(\tau)&=\frac{3}{4}(\mu_a^2I_1+I_2)e^{-\tau/\mu_a}+\\&\frac{3}{4}\int_0^\infty\int_0^1\frac{e^{-|\tau-\tau'|/\mu'}}{2\mu'}\mu'^2[-J_2(z)([\mu'^2(G-iH)]+[\mu'^2(G+iH)])+iJ_1(z)([-\mu'v'(-iD-E)]+[\mu'v'(iD-E)])+J_0(\\&+\frac{3}{4}\int_0^\infty\int_0^1\frac{e^{-|\tau-\tau'|/\mu'}}{2\mu'}[-J_2(z)([-(G-iH)]+[-(G+iH)])+iJ_1(z)((0)-(0))+J_0(z)A]\partial\mu'\partial\tau'\end{aligned} \quad (76)$$

Simplifying, we get:

$$\begin{aligned}A(\tau)&=\frac{3}{4}(\mu_a^2I_1+I_2)e^{-\tau/\mu_a}\\&+\frac{3}{4}\int_0^\infty\int_0^1\frac{e^{-|\tau-\tau'|/\mu'}}{2\mu'}\mu'^2[-J_2(z)(2[\mu'^2(G)])+iJ_1(z)([2\mu'v'(iD)])+J_0(z)[A+B(1-\mu'^2)]]\partial\\&+\frac{3}{4}\int_0^\infty\int_0^1\frac{e^{-|\tau-\tau'|/\mu'}}{2\mu'}[-J_2(z)([-2(G)])+J_0(z)A]\partial\mu'\partial\tau'\\&=\frac{3}{4}(\mu_a^2I_1+I_2)e^{-\tau/\mu_a}\\&+\frac{3}{4}\int_0^\infty\int_0^1\frac{e^{-|\tau-\tau'|/\mu'}}{2\mu'}J_2(z)(2(1-\mu'^4))G-J_1(z)(2\mu'^3v'D)+J_0(z)[(\mu'^2+1)A+\mu'^2B(1-\mu'^2)\end{aligned} \quad (77)$$

We simplify the G coefficients. Using equation (2) we get:

$$\begin{aligned}(1-\mu^4)&=1-\mu^2(1-v)^2\\&=1-\mu^2+\mu^2v^2\\&=v^2+\mu^2v^2\end{aligned} \quad (78)$$

So that using equations (21) and (22) we simplify (77):

$$\begin{aligned}A(\tau)&=\frac{3}{4}(\mu_a^2I_1+I_2)e^{-\tau/\mu_a}\\&+\frac{3}{4}[2(M_{20}+M_{24})*G-2M_{13}*D+(M_{00}+M_{02})*A+(M_{02}-M_{04})*B]\end{aligned} \quad (79)$$

We set $\mu_a^2=1$ because the viewing geometry is in the direct backscattering so $\mu_a=1$:

$$A=\frac{3}{4}((I_1+I_2)e^{-\tau}+(M_{00}+M_{02})*A+(M_{02}-M_{04})*B-2M_{13}*D+2(M_{20}+M_{22})*G) \quad (80)$$

Which is the top line of equation (20), as required.